# Real-time Operation Optimization of Microgrids with Battery Energy Storage System: A Tube-based Model Predictive Control Approach

Cheng Lyu, *Student Member, IEEE*, Youwei Jia, *Member, IEEE*, and Zhao Xu, *Senior Member, IEEE*

*Abstract*—Battery energy storage systems (ESS) are widely used in microgrids to complement high renewables. However, the real-time energy management of microgrids with battery ESS is challenging in two aspects: 1) the evolution process of battery energy level is across-time coupled; 2) uncertainties unavoidably arise in the forecasting process for renewable generation. In this paper, a tube-based model predictive control (MPC) approach is innovatively proposed in accommodating the real-time energy management of microgrids with battery ESS. Firstly, a real-time operation model of battery, including the degradation cost and time-aware SoC range, is proposed for the battery ESS. In particular, the battery feature *shallower-cheaper* is depicted and the terminal SoC requirement is achieved. Secondly, two cascaded MPC controllers are designed in the proposed tube-based MPC, in which reference trajectories are generated by the nominal MPC without uncertainties, and then the ancillary MPC steers the actual trajectories to the nominal ones upon the realization of uncertainties. Specifically, in this paper, the battery SoC is viewed as the state variable of the system, while the generator power output and exchange power with the utility are seen as control variables. Lastly, numerous case studies demonstrate the effectiveness of the proposed approach, including both the low and high penetration level of renewables. Additional Monte Carlo simulations of consecutive 365 days show that the competitive ratio of the proposed approach is excellently below 1.10.

*Index Terms*--Real time operation, tube-based model predictive control, battery energy storage system.

## I. INTRODUCTION

OVER the last decades, intermittent renewables, such as wind energy and solar photovoltaic (PV) energy, have been widely exploited to resolve the carbon emission issues in microgrids. As an emerging paradigm of future smart grids, microgrids provide a promising solution to interconnect various distributed renewable energies (DRE), energy storage system (ESS), and dispatchable generators [1-3].

Similar to conventional power systems, the real-time operation optimization of microgrids aims to securely minimize the operational cost, while subject to real-time power supply-demand balance [4]. Nevertheless, there are two main challenges confronted by the real-time microgrid energy management. Firstly, the uncertain nature of DRE output complicates the design of energy management framework. There are inevitably errors in predicting the wind speed or solar irradiation. As such, the accumulated forecast errors inhibit the economy of the day-ahead energy schedule plan. Secondly, it is difficult to tackle the real-time degradation of battery ESS given its healthy operational requirements. As such, it is also critical to investigate the real-time operation optimization strategies for battery ESS embedded microgrids.

To tackle the first challenge, several efforts have been made in recent years to address the uncertainty issues of energy management. Firstly, stochastic programming (SP) generates representative scenarios of DRE output. In [5], for example, a two-stage stochastic problem is formulated to capture the power system operation, in which the first-stage problem determines the optimal dispatch of generators, while the second-stage problem takes the uncertainty into account. Though SP provides a tractable method in dealing with the uncertainty, the scenario generation mechanism requires a known distribution of parameters and suffers from computation burdens. Secondly, [6] adopts robust optimization (RO) for the energy management of power system, in which the worst distributions in the defined uncertainty set are considered. As such, RO can reduce the computation burden but leads to a relatively conservative solution, as compared to SP [7]. To bridge the gap between SP and RO, distributionally robust optimization (DRO) arises to embrace limited distribution information into RO, e.g., means and variances [8]. Despite its general potential in handling uncertainty, the handling of high dimensionality in conic optimization or semi-definite optimization seems to prevent its direct application in practice [9, 10].

To this end, handling uncertainty explicitly, MPC approach is regarded as a feasible and computationally tractable alternative to aforementioned methods [11]. MPC has drawn attention in the energy management field due to three factors: 1) it considers predictions and future behaviors of the system, which is desirable for the power system operation that greatly rely on forecasts; 2) it designs a feedback mechanism, which inherently provides robustness against the uncertainty; and 3) it can deal with power system constraints, such as technical limit

Cheng Lyu is with the Department of Electrical Engineering, The Hong Kong Polytechnic University, Hong Kong. (e-mail: cheng.lyu@connect.polyu.hk)
Youwei Jia is with the Department of Electrical and Electronic Engineering, University Key Laboratory of Advanced Wireless Communication of Guangdong Province, Southern University of Science and Technology, 518088, Shenzhen, China. (e-mail: jiayw@sustech.edu.cn)
Zhao Xu is with Shenzhen Research Institute and Department of Electrical Engineering, The Hong Kong Polytechnic University, Hong Kong. (e-mail: eezhaoxu@polyu.edu.hk)

constraints, and battery SoC constraints. MPC is an optimal control method to deliver the control sequence by solving a constrained optimization problem in a finite time horizon [11]. Recently, tube-based MPC is emerging as an advanced version of MPC to better address the uncertainty issues, in which two cascaded MPC controllers are devised [12]. In particular, a nominal MPC without uncertainties generates the central trajectory as a reference of the system; and then an ancillary MPC steers the actual trajectory of the uncertain system to the nominal trajectory [13]. However, the application potential of tube-based MPC in ESS-embedded microgrids is still unclear.

To overcome the second challenge, several research attempts have been reported in the literatures, regarding the degradation cost model of battery ESS, especially lithium-ion batteries due to its high energy density and long lifespan [14]. As in most existing literatures, state of charge (SoC) plays a key role in the economic operation of battery ESS [14]. On the one hand, SoC level represents the health condition of the battery. That is, the battery cannot be fully discharged, or be charged to a full storage level. On the other, the SoC evolution is a time-coupled process, which records the historical charging and discharging profiles. However, the time-coupled SoC evolution process inhibits decoupling the energy management into every single time slots. In addition, the degradation cost of battery is another key issue in real-time operation, towards which most previous works consider a constant cost per charging or discharging energy for the sake of simplicity [5, 15]. In this regard, recent work [16] shows that the battery capacity degrades with calendar aging and cycle aging of lithium ion battery cells. In [17], the offline rain-flow cycle counting algorithm is applied to the cycle aging assessment of batteries. In [18], the authors provide a practical way to approximate the degradation cost, in which the degradation cost only occurs at the discharging half cycle. However, the cycle depth of batteries is not known to the system operator until the end of the discharging cycle, which makes it difficult to calculate the cycle aging cost online.

To fill in the research gap, this paper newly proposes a novel approach for the real-time energy management of ESS-embedded microgrids. Firstly, a real-time operation model is proposed for battery ESS, which includes the degradation cost and time-aware SoC operational range. Secondly, two cascaded MPC are specially designed in the proposed tube-based MPC, in which reference trajectories are generated by the nominal MPC without uncertainties, while the bound constraints are tightened. Subsequently, the ancillary MPC steers the actual trajectories to the nominal ones by solving the constrained optimization problem with original constraints, after the realization of uncertainties. In particular, the battery SoC is viewed as the state variable of the system, while the generator power output and exchange power with the utility are seen as control variables. Lastly, numerous case studies and 1-year Monte Carlo simulations demonstrate the extraordinary performance guarantee of the proposed approach.

The rest of the paper is organized as follows: Section II describes the real-time operation model and problem without uncertainties. Section III presents basic concepts of tube-based MPC as a preliminary. In section IV, the real-time operation problem based on tube-based MPC is formulated. Section V validates the proposed approach by case studies. Section VI summarizes our work.

## II. REAL-TIME OPERATION MODEL AND PROBLEM

In this section, the real-time degradation of battery ESS is analyzed and then the energy management problem without uncertainty are formulated.

### A. Battery Degradation

Batteries have limited cycle life owing to the fading of active materials caused by charging and discharging cycles. The most significant part of the battery degradation cost is the operating expense caused by cycling aging. Fig. 1 shows an empirical curve of cycle depth aging for lithium-ion batteries. As is seen, the marginal degradation cost is not a fixed value. Specifically, the life fading effect is less serious when the battery discharges to a smaller cycle depth. In other words, shallow charging and discharging cycles incur small damages to the battery, as compared to deeper ones. Under this condition, the lifetime of batteries can no longer be viewed as a fixed value, and the unit cost of battery energy is no longer treated as a constant expense [18].

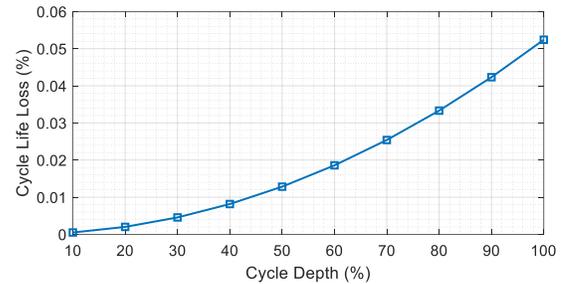

Fig. 1. Cycle life loss curve of battery ESS.

### B. Degradation Cost Model

Suppose the cycle life loss curve for lithium-ion batteries, as in Fig. 1, is characterized by the following function:
$$\varphi(\theta_t)=\alpha\theta_t^{1+\beta}, \qquad \theta_t \in [0, 1] \qquad (1)$$
where $\varphi$ represents the cycling life loss, $\theta_t$ represents the discharging depth at time slot $t$, and parameters $\alpha, \beta \geq 0$ implies the curve shape. It is worth noting that only the discharging depth is used here to represent the charging and discharging cycle, because of that a complete cycle depth contains two equal half-cycles: the charging half-cycle and the discharging half-cycle. As such, the marginal cost of a battery is derived by taking the derivative of (1) with respect to the discharging power [18]:
$$\frac{\partial \varphi(\theta_t)}{\partial d_t}=\frac{\mathrm{d}\varphi(\theta_t)}{\mathrm{d}\theta_t}\frac{\partial \theta_t}{\partial d_t}=\frac{1}{\eta^d B}\frac{\mathrm{d}\varphi(\theta_t)}{\mathrm{d}\theta_t} \qquad (2)$$
where the second equality holds because of the discharging cycle depth evolution equation:
$$\eta^d B(\theta_t - \theta_{t-1})=d_t \qquad (3)$$
where $B$ is the capacity of the battery, $\eta^d$ is the discharging efficiency coefficient, and $d_t$ represents the discharging power at time slot $t$.

Inspired by the piecewise method in [18], an *N*-segmental degradation cost model is adopted in this paper to approximate the real-time degradation cost as in (4):

$$C_i = \frac{R}{\eta^d B} \frac{\varphi(\frac{i}{N}) - \varphi(\frac{i-1}{N})}{1/N}, \quad \theta_t \in \left[\frac{i-1}{N}, \frac{i}{N}\right) \quad (4)$$

where $C_i$ denotes the degradation cost of *i*-th segment, $R$ is the replacement expense of the battery ESS, the segment index $i$ =1, 2, …, $N$ and $N$ is the number of segments.

Two remarks are delivered to give more explanations regarding the segmental degradation cost model:

1) The proposed degradation cost model (4) captures the marginal damage of per discharging power on the cycle life loss. Instead of utilizing a fixed cost like most existing works, the proposed segmental model can reflect the operating feature - "the shallower, the cheaper" of the battery.

2) Fig. 2 briefly illustrates the proposed segmental model with *N*=5. As can be observed, the total capacity of the battery is equally divided into 5 segments, with index 1 to 5, and each segment shares 1/5 of the total capacity. More importantly, the degradation cost of five segments are $C_1 < C_2 < C_3 < C_4 < C_5$, calculated by (4). In the real-time operation, the shallower segments have priority to charge and discharge.

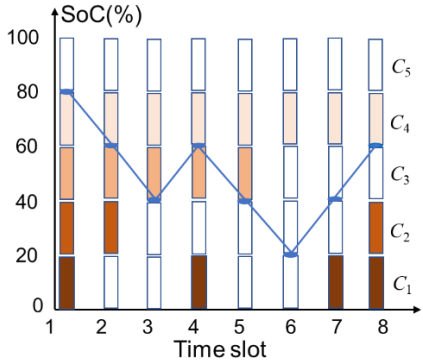

Fig. 2. Operation illustration of the proposed segmenal cost model for battery ESS with *N*=5.

### C. Time-aware SoC Operational Range

SoC is the critical state variable to track the operation of battery ESS, of which the evolution process can be formulated as follows:

$$S_t = S_{t-1} + \frac{\eta^c c_t \tau}{B} - \frac{d_t \tau}{\eta^d B} \quad (5)$$

where $S_t, c_t$ are SoC level, charging power of the battery at time slot $t$, $\eta^c$ is charging efficiency coefficient, and $\tau$ is the time slot interval, respectively. The operation of battery should subject to technical requirements such as the SoC range, charging and discharging power limits, due to the health requirements of battery:

$$S_{\min} \leq S_t \leq S_{\max} \quad (6)$$

$$0 \leq c_t \leq c_{\max}(1 - u_t) \quad (7a)$$

$$0 \leq d_t \leq d_{\max} u_t \quad (7b)$$

where $S_{\min}$ and $S_{\max}$ indicate the minimal and maximal SoC of battery; $c_{\max}$ and $d_{\max}$ indicate the maximal charging and discharging power of battery; the binary variable $u_t$ is introduced to avoid the occurrence of simultaneous charging and discharging in a single period.

In addition, daily operation of battery ESS is continuous inter days, which means that the terminal SoC of the scheduled day becomes the initial SoC of the next day. To capture this inter-day characteristic, equation (8) requires that the terminal SoC should be within a prescribed interval.

$$S_{end1} \leq S_T \leq S_{end2} \quad (8)$$

However, it should be noticed that (8) not only regulates the SoC range constraint at the terminal time slot *T*, but also exerts underlying impacts for several periods ahead of *T*. This is because of the limited battery charging and discharging power in (7). To this end, we propose a time-aware SoC boundary constraint as follows:

$$S_{t,\min} \leq S_t \leq S_{t,\max} \quad (9a)$$

$$S_{t,\min} = \max\{S_{\min}, S_{t+1,\min} - \eta^c c_{\max}\tau/B\} \quad (9b)$$

$$S_{t,\max} = \min\{S_{\max}, S_{t+1,\max} + d_{\max}\tau/\eta^d B\} \quad (9b)$$

It is worthwhile noting that (9a) differs from (6) in that the SoC boundary values are no longer fixed. Instead, the proposed time-aware boundaries are calculated by (9b) and (9c) in the backward order of *t*=*T*-1, *T*-2, *T*-3, …,1.

Moreover, to coincide with the segmental cost model, the following constraints are added to capture the relationship between variables of all segments and the whole battery:

$$S_t = \sum_{i=1}^{N} S_{i,t}, \quad c_t = \sum_{i=1}^{N} c_{i,t}, \quad d_t = \sum_{i=1}^{N} d_{i,t} \quad (10)$$

where $S_{i,t}$ denotes the state of charge of *i*-th segment among all $N$ segments of battery ESS, $c_{i,t}$ and $d_{i,t}$ are charging and discharging power of *i*-th segment at time slot *t*, respectively. Hence, similar to (5), the energy evolution of each segment should follows, for each segment *i*:

$$S_{i,t} = S_{i,t-1} + \frac{N\eta^c \tau}{B} c_{i,t} - \frac{N\tau}{\eta^d B} d_{i,t} \quad (11)$$

### D. Real-time Operation Cost and Constraints

The cost function of operation optimization in microgrids includes the trading cost with utility, fuel cost of dispatchable generators and cycle degradation cost of battery ESS. A quadratic cost function is used to model the fuel consumption cost of local generators. Accordingly, the real-time operation cost is given as follows:

$$F_t = \sum_{j \in \mathcal{G}} \left(a_{2,j} g_{j,t}^2 + a_{1,j} g_{j,t} + a_{0,j}\right) + b_t p_t^{buy} - s_t p_t^{sell}$$
$$+ \sum_{i=1}^{N} \left(C_i d_{i,t} + \varepsilon C_i c_{i,t}\right) \quad (12)$$

where $a_{2,j}$, $a_{1,j}$ and $a_{0,j}$ are cost parameters for dispatchable generators *j*; $g_{j,t}$ denotes the power generation of generator *j*; $\mathcal{G}$ represents the generator set; $b_t$ and $c_t$ denote the power purchased from and sold to the utility; the corresponding prices are $p_t^{buy}$ and $p_t^{sell}$ with $p_t^{buy} > p_t^{sell}$; $\varepsilon$ is the parameter associated with the charging cost, respectively. Note that $\varepsilon$ is set small enough for two reasons: 1) the degradation model is derived based on the discharging half-cycle, so in principle the charging half-cycle is no longer necessary in the cost function; 2) the positive $\varepsilon$ is introduced as a signal to give charging priority to shallower segments.

The real-time power supply-demand balance and relative technical constraints are shown as follows:

$$b_t - s_t + d_t - c_t + r_t + \sum_{j \in \mathcal{G}} g_{j,t} = (1+\theta)l_t \quad (13)$$

$$0 \leq b_t \leq b_{\max} v_t \quad (14a)$$

$$0 \leq s_t \leq s_{\max}(1 - v_t) \quad (14b)$$

$$g_{j,\min} \leq g_{j,t} \leq g_{j,\max} \quad (15a)$$

$$g_{j,t} - g_{j,t-1} \leq g_j^{RU} \quad (15b)$$

$$g_{j,t-1} - g_{j,t} \leq g_j^{RD} \quad (15c)$$

where (13) represents the power supply-demand balance, in which $r_t$ and $l_t$ denote the forecast values of DRE generation and load demand, the parameter $\theta$ is introduced to represent power distribution loss; (14) depicts the constraint of power exchange with the utility, in which the binary variable $v_t$ is introduced to suggest the power exchange direction; (15a) represents the power output limit of generators; (15b) and (15c) are ramping capacity constraints, in which $g_j^{RU}$ and $g_j^{RD}$ are ramping-up and ramping-down rates of generator $j$.

## III. TUBE-BASED MPC APPROACH

In this section, some basic concepts are presented as preliminaries for the tube-based MPC approach.

### A. Overview of Tube-based MPC

It is noted that the problem described in section II does not taken into uncertainty into account. Tube-based MPC is naturally a computationally tractable approach to deal with the system under uncertainty. Generally, we consider an uncertain system modeled as follows:

$$x_{t+1} = Ax_t + Bu_t + \Delta w_t \quad (16)$$

where $x$ represents the state of the system, $u$ represents the control action, and $\Delta w$ represents the uncertainty, respectively. Both state and control variables are subject to the constraints as follows:

$$x_t \in \mathbb{X}_t, \qquad u_t \in \mathbb{U}_t \quad (17)$$

As shown in Fig. 3, the tube-based MPC consists of two cascaded MPC controllers: a nominal MPC controller that generates a central trajectory subject to tightened bound constraints; an ancillary MPC controller that steers the trajectory of an uncertain system to the nominal trajectory.

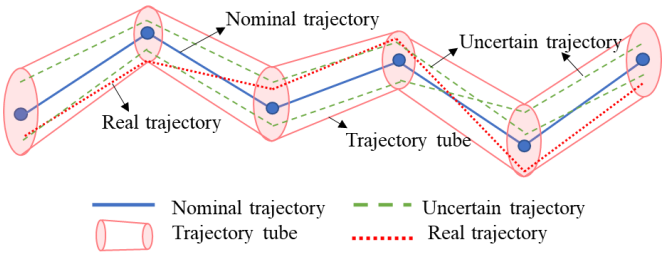

Fig. 3. Illustration of trajectories of tube-based MPC.

### B. Nominal MPC

The nominal system is obtained from (16) by neglecting the uncertainty term $\Delta w$:

$$\bar{x}_{t+1} = A\bar{x}_t + B\bar{u}_t \quad (18)$$

where $\bar{x}$ denotes the nominal system state and $\bar{u}$ denotes the nominal control, respectively. Both state and control sequences are subject to the tightened constraints as follows:

$$\bar{x}_t \in (1-\rho_x)\mathbb{X}_t, \qquad \bar{u}_t \in (1-\rho_u)\mathbb{U}_t \quad (19)$$

where two parameters $\rho_x, \rho_u \in (0,1)$ are introduced to define the tightened bound constraints. By properly choosing these two parameters, the robustness of the control will be guaranteed. On the one hand, they are expected to be smaller to endure the uncertain disturbance $\Delta w$; on the other hand, they should be as large as possible to avoid over-conservative results. The objective function and remaining constraints of the nominal MPC are the same with the original problem without uncertainty.

### C. Ancillary MPC

The task of the ancillary controller is to keep the state and control sequence of the uncertain system into a tube centered on the nominal trajectory determined previously. The ancillary MPC achieves this target by solving an optimal control problem that minimizes the cost of deviation of the state $x$ and control $u$ from the nominal trajectories:

$$\min \sum_{N_x} \mu_x (x_t - \bar{x}_t)^2 + \sum_{N_u} \mu_u (u_t - \bar{u}_t)^2 \quad (20)$$

where $N_x$ and $N_u$ denote the number of state and control, $\mu_x$ and $\mu_u$ are weights associated with the deviation of state and control variables, respectively. The constraints in the ancillary MPC are the same as the original problem, except that the uncertainty term $\Delta w$ is taken into account.

## IV. TUBE-BASED MPC PROBLEM FORMULATION

In this section, the real-time microgrid operation optimization problem with uncertainty is formulated based on tube-MPC approach.

### A. Nominal MPC Design

The nominal system is obtained by neglecting the uncertainty of DRE generation and load demand. That is, the forecast values are assumed to be accurate in the nominal MPC. Bearing this in mind, the objective function of nominal MPC is to minimize the total cost over the prediction time horizon $\mathcal{H}_1$ as follows:

$$\min \sum_{t \in \mathcal{H}_1} F_t \quad (21)$$

where $F_t$ is defined in (12). As discussed in section III, some of constraints are tightened in the nominal MPC. In this paper, the battery SoC range constraints (9) and power generation constraint (15a) of dispatchable generators are tightened as follows:

$$(1+\rho_1)S_{t,\min} \leq S_t \leq (1-\rho_1)S_{t,\max} \quad (22)$$

$$(1+\rho_2)g_{j,\min} \leq g_{j,t} \leq (1-\rho_2)g_{j,\max} \quad (23)$$

where parameter $\rho_1, \rho_2 \in (0,1)$ are introduced to tighten the constraint. To ensure the robustness, they should be chosen carefully. In this paper, we tune the parameters $\rho_1$ and $\rho_2$ by offline Monte Carlo simulations to achieve the balance between feasibility and economy. The remaining real-time operation constraints of the nominal MPC are the same as discussed above.

*B. Ancillary MPC Design*

In the real-time operation, the instant energetic imbalance should be coped with upon the realization of uncertainty. That is, the real-time power supply-demand balance is as follows:

$$b_t - s_t + d_t - c_t + r_t + \Delta r_t + \sum_{j \in \mathcal{G}} g_{j,t} = (1+\theta)(l_t + \Delta l_t) \quad (24)$$

where $r_t + \Delta r_t$ and $l_t + \Delta l_t$ denote the actual GRE output and load demand at time slot $t$.

The ancillary MPC aims to keeps the state and control variables of the uncertain system into a tube centered on the nominal trajectory determined previously. Specifically, in this paper, the battery SoC is viewed as the state variable of the system, while the generator power output and power trading amount are viewed as the control variables. Hence, the objective function of the ancillary MPC is as follows:

$$\min \sum_{t \in \mathcal{H}_2} \mu_x \left(S_t - S_t^{nom}\right)^2 + \mu_u \left[ \begin{array}{l} \left(b_t - b_t^{nom}\right)^2 + \left(s_t - s_t^{nom}\right)^2 \\ + \sum_{j \in \mathcal{G}} \left(g_{j,t} - g_{j,t}^{nom}\right)^2 \end{array} \right] \quad (25)$$

where $S_t^{nom}, b_t^{nom}, s_t^{nom}, g_{j,t}^{nom}$ are nominal trajectories of battery SoC, purchasing power, selling power and power generation of generator $j$, resulting from the nominal MPC controller; $\mu_x$ and $\mu_u$ are weights of SoC part and power part; $\mathcal{H}_2$ denotes the time horizon of ancillary MPC. In practice, the length of $\mathcal{H}_2$ is less than that of $\mathcal{H}_1$. The constraints of the ancillary MPC are the same as the original problem as in Section II, except that the power balance (13) is updated to (25).

At each time slot, a resultant optimal dispatch sequence is obtained, of which the first element is applied to the system. Then the horizon window is rolling forward. At the next time period, the nominal and ancillary MPC are repeated using the newly measured states and forecasting information. By doing so, a feedback is designed in the receding horizon manner.

V. CASE STUDY

In this section, numerical tests are conducted to validate the proposed model and approach for a 24-hour real-time operation. The formulated mixed integer problem is solved by the solver 'GUROBI' on MATLAB 2019b.

*A. Microgrid Configuration*

The considered microgrid contains two dispachable generators, one battery ESS, and rooftop solar PV panels. Table I provides parameters of the battery ESS, and the cycle life loss function parameters are $\alpha = 5.24 \times 10^{-4}$, $\beta = 1.03$, respectively. Table II shows the parameters of the dispatchable generators. The maximal power exchange with the utility is set as $s_{max} = b_{max} = 250$ kW, due to power line distribution limit.

TABLE I. BATTERY ESS DATA

| $B$ (kWh) | $R$ ($) | SoC range | Terminal SoC | $c_{max}/d_{max}$ | $N$ | $\eta^c/\eta^d$ |
|---|---|---|---|---|---|---|
| 400 | 80000 | 20%~90% | 50%~60% | 80kW | 10 | 0.95 |

TABLE II. GENERATOR PARAMETERS

| Gen | $a_2$ ($/(kW)^2$h) | $a_1$ ($/kWh) | $a_0$ ($) |
|---|---|---|---|
| #1 | 0.0013 | 0.062 | 0 |
| #2 | 0.0010 | 0.057 | 0 |
| Gen | $g^{RU(RD)}$ (kW) | $g_{max}$ (kW) | $g_{min}$ (kW) |
| #1 | 240 | 52 | 6 |
| #2 | 280 | 92 | 16.4 |

The power buying price from the utility is 0.116 $/kWh during peak periods (7:00-21:00) and 0.072 $/kWh during off-peak periods. The power selling price is assumed to be half of the selling price. The time interval is 15 minutes. In the tube-based MPC, the horizon length of $\mathcal{H}_1$ and $\mathcal{H}_2$ are chosen as 8 and 2 respectively. That is, the 2-hour forecast information is used to generate a nominal trajectory, while only 30-minite data is used to determine the actual control sequences. Some key parameters of the proposed approach are shown in Table III.

TABLE III. PARAMETERS

| $\rho_1$ | $\rho_2$ | $\theta$ | $\varepsilon$ | $\mu_x$ | $\mu_u$ |
|---|---|---|---|---|---|
| 0.05 | 0.1 | 0.01 | 0.001 | 400 | 1 |

*B. Case 1: Low Penetration of DRE*

In case 1, the microgrid is supposed to install a low penetration of DRE. Fig. 4 illustrates the forecast and actual profiles of DRE generation and load demand of the day, in which the forecast error of DRE generation is larger than that of load demand. As can be observed, the forecast output of solar energy is less than the load demand.

Fig. 5 illustrates the segmental cost with $N=10$ and proposed time-aware SoC range. The total capacity of battery ESS is equally divided into 10 segments, each of which has the sub capacity 40kWh. In Fig. 5(a), the segmental degradation cost increases with the segment index. This feature depicts that a deep discharging cycle incurs a high degradation cost in batteries, as stated in section II-B. Moreover, since the life cycle loss function in (1) is quasi-quadratic, the cost curve in Fig. 5(a) seems linear with respect to the segment index. In Fig. 5(b), the time-aware SoC boundary is stricter than the original SoC boundary, [20%, 90%], of which the difference is illustrated as shaded area. For example, the time-aware SoC range becomes half of the original SoC range at 23:30. In fact, the shaded area suggests the resulting dead region of battery with terminal SoC requirement [50%, 60%]. That is, the proposed time-aware SoC is essentially the necessary condition for the terminal SoC requirement, owing to limited charging and discharging power.

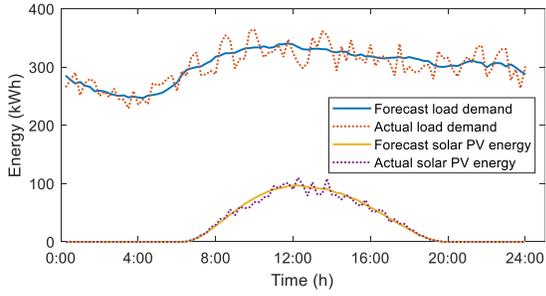

Fig. 4. Forecast and actual profiles of solar energy and load demand in case 1.

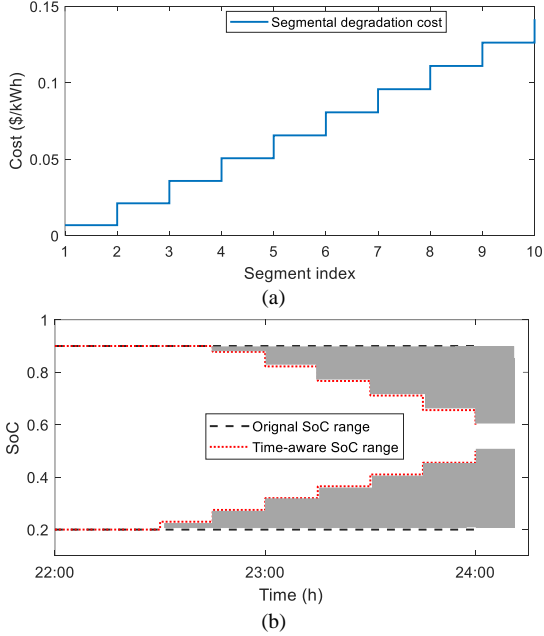

Fig. 5. Real-time operation model of battery ESS. (a) Segmental degradation cost model; (b) Time-aware SoC range.

Fig. 6 illustrates the dispatch results of the proposed approach. In Fig. 6(a), the SoC profiles of the battery ESS are shown with the initial SoC = 60%, and terminal SoC = [50%, 60%]. As is observed, the actual SoC trajectory fluctuates along with the nominal one, upon the realization of uncertainty. This is because of the fact that the ancillary MPC minimizes the deviation between actual trajectories and nominal trajectories. This mechanism allows the actual SoC trajectory lie within the 'tube' centered around the nominal SoC trajectory.

The power generation profiles and power trading profiles in the nominal MPC and ancillary MPC are illustrated in Fig.6(b) and Fig. 6(c), respectively. As the figures show, the power output and trading are constrained by their maximal and minimal limit. For example, at periods 11:00-11:30, the actual dispatch power of generators is more than the nominal value. This is attributed to the tightened constraint feature in the tube-based MPC, which shows the robustness advantage over conventional MPC. In addition, the actual value also fluctuates along the nominal trajectory to response to the real-time solar power output and load demand. Generally, the real-time energetic imbalance is mostly complemented by local generators instead of exchanging with the utility due to the high purchasing price. These above-mentioned results indicate the foresight of tube-MPC that leads to a solution robust to forecast errors.

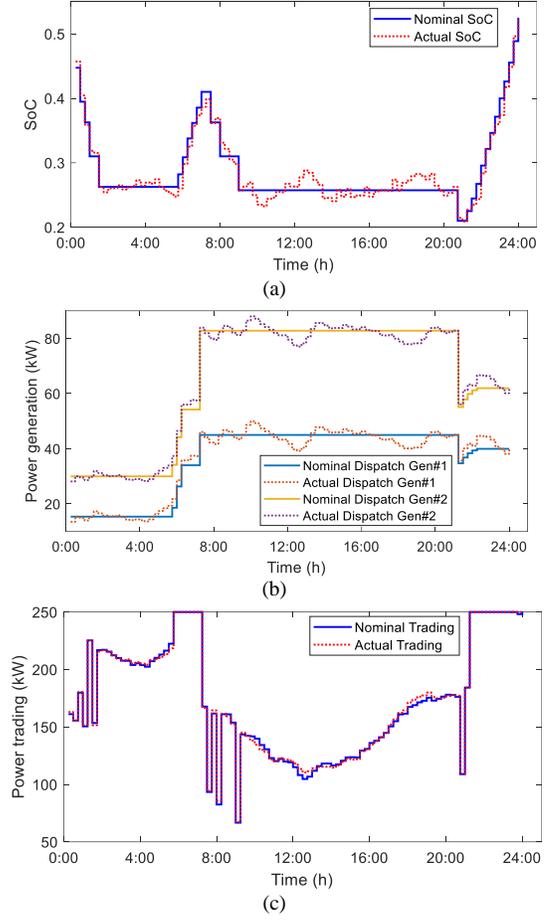

Fig. 6. Dispatch results of the proposed tube-based MPC in case 1. (a) SoC trajectory of the battery ESS; (b) Power dispatch of dispatchable generators; (c) Power trading with the utility.

### C. Case 2: High Penetration of DRE

In case 2, we are aimed to investigate the performance of tube-based MPC approach in microgrids with high penetration of DRE. In this case, Fig. 7 shows the solar energy output and load demand profiles, of which the energy levels are comparable.

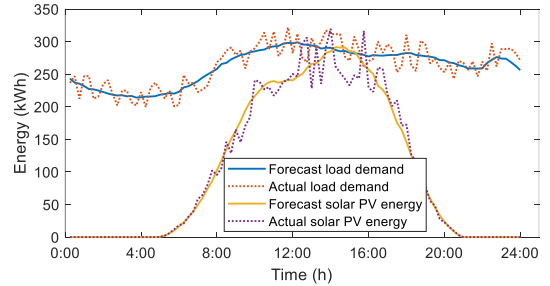

Fig. 7. Forecast and actual profiles of solar energy and load demand in case 2.

The battery SoC trajectories, the power generation and trading profiles are shown in Fig. 8. As is shown, the actual profiles are all bounded by the maximal and minimal limit, and the feasibility of the system is guaranteed. In addition, we can see from Fig. 8(a) that the actual SoC boundaries touches the minimal value 20% during some time slots in 8:00-16:00. For example, at 12:00, the nominal SoC is 21% instead of the minimal SoC 20%, where the 1% margin is allowed for the uncertainty. As a matter of fact, the actual SoC at this time slot is 20.01 %, which means that the battery ESS discharges more than the nominal plan. This observation also corroborates the foresight of tube-based MPC, in which the nominal MPC tightens the SoC constraints to guarantee the feasibility and robustness, though the uncertainty is unknown beforehand.

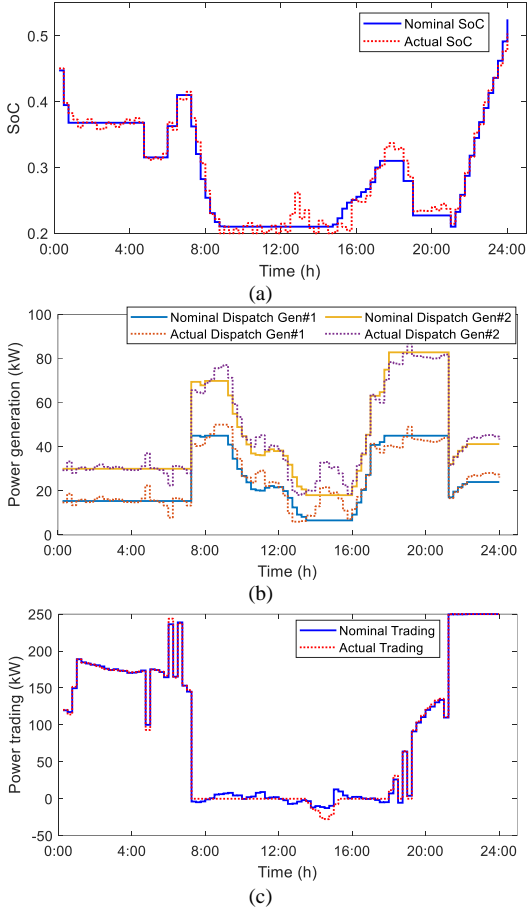

Fig. 8. Dispatch results of the proposed tube-based MPC in case 2. (a) SoC trajectory of the battery ESS; (b) Power dispatch of dispatchable generators; (c) Power trading with the utility.

### D. Economic Evaluation of Tube-based MPC

To evaluate the economy performance of the tube-based MPC approach, the metric named 'competitive ratio' is adopted here, which is defined as the approximation ratio achieved by the designed approach compared to the ideal offline results [4]. As is shown in Table IV, for case 1, the total daily cost is $574.39 with the proposed tube-based MPC approach, and the offline perfect dispatch cost is $554.25. Accordingly, the competitive ratio of proposed approach is 1.0363. Note that the perfect dispatch cost is not known until the end of the day. Similar results can be found in the case 2, in which the competitive ration is 1.0588.

TABLE IV.  OPERATION COST RESULT

| Metric | Value in Case 1 | Value in Case 2 |
|---|---|---|
| Perfect dispatch cost by offline optimization ($) | 554.25 | 277.55 |
| Total cost calculated by tube-based MPC ($) | 574.39 | 293.87 |
| Cost increase amount ($) | 20.14 | 16.32 |
| Competitive ratio | 1.0363 | 1.0588 |

Furthermore, Monte Carlo simulations are conducted for 365 consecutive days based on the realistic electricity data [19, 20], as shown in Fig. 9. The uncertainties of DRE output and load demand forecasting values are set as 20% and 10%, respectively. By applying the proposed approach for the real-time energy management, the competitive ratios that can be achieved are plotted in Fig. 10. The maximal competitive ratio is 1.086 on the day 138, while the minimal one is just 1.028 on the day 331. One interesting notice is that the metric is larger in summer than that in winter, which may be the consequence of peak demands in summer seasons. These simulation results show the advantages of tube-based MPC approach over existing online algorithms with the competitive ratio ranging from 1.1 to 1.6, as reported in [21].

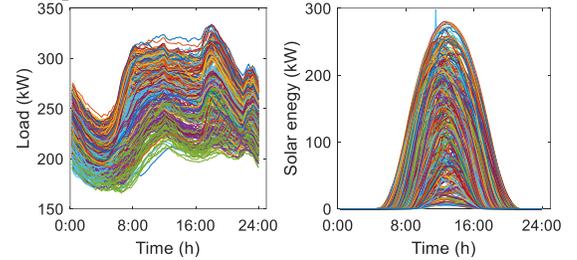

Fig. 9. Forcast value of load demand and soalr energy in consecutive 365 days.

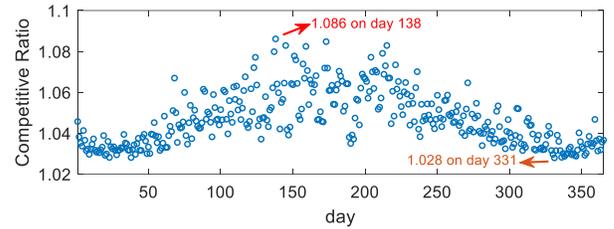

Fig. 10. Competitive ratio of cases in consecutive 365 days.

### E. Influence of Constraint Tightening Parameters

In the tightened constraint (22) and (23), parameters $\rho_1$ and $\rho_2$ are regarded as the robustness parameters that control the balance between economy and robustness. In previous cases, they are chosen as 0.05 and 0.1 by offline Monte Carlo simulations. In this subsection, we study their influence on the approach performance.

The case of day 138 is chosen as an example, since it is the worst case as discussed above. Fig. 11 shows the competitive ratio results for various $\rho_1$ and $\rho_2$. It is seen that the competitive ratio is strictly below 1.10 in all circumstances regardless of $\rho_1$ ranging from 0.01 to 0.09, and $\rho_2$ ranging from 0.5 to 0.2. The results once again validate the excellent performance of tube-based MPC, which is key to future practical application.

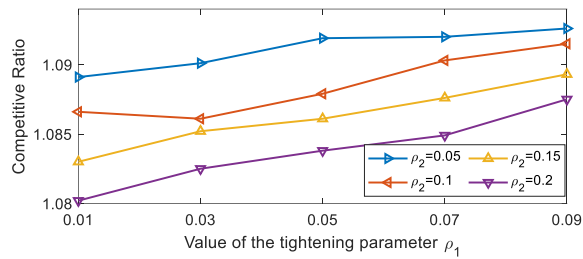

Fig. 11. Influence of Constraint Tightening Parameters.

*F. Influence of Horizon Length*

It is noted that the length of horizon window may affect the performance of the proposed approach, which is optimized on the truncated time window. Fig. 12 shows the competitive ratio of the consecutive 365 days with varying horizon length of the ancillary MPC from 1 to 4. Since the dispatch interval is 15min/slot, the corresponding horizon is 15min to 1 hour. As seen from the figure, the competitive ratios are generally below 1.10, while the middle value is between 1.04 and 1.05. It is noticed that outliers disappear in the cases where the horizon length extends to 3 and 4, which means that the results are more concentrated and the worst performance is improved by increased horizon length.

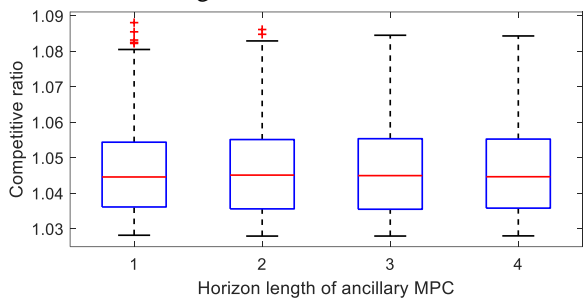

Fig. 12. Influence of horizon length.

## VI. CONCLUSION

In this paper, a tube-based MPC approach is innovatively proposed for the real-time energy management of microgrids with battery ESS. Two cascaded MPC controllers are designed in the proposed tube-based MPC, in which the nominal MPC generates the control and state trajectories as a reference by tightening some constraints, and the ancillary MPC steers the actual trajectory to the nominal trajectory upon the realization of the uncertainty. Specifically, in this paper, the battery SoC is viewed as the state variable of the system, while the generator power output, and utility power trading amount are viewed as the control variables. In addition, an intact real-time operation model is newly proposed for battery ESS to participate the real-time operation. Numerous cases demonstrate the effectiveness of the proposed approach. Monte Carlo simulations demonstrated the performance guarantee of the proposed approach, of which the competitive ratio is excellently below 1.10. Future correlative works include the extensive application in multiple ESS-embedded system.